\documentclass[superscriptaddress,twocolumn,showpacs,amsmath,amssymb,nofootinbib]{revtex4}
\usepackage{graphicx}
\usepackage{dcolumn}
\usepackage{bm}

\begin{document}

\title{One-body energy dissipation in fusion reaction 
from mean-field theory}

\author{Kouhei Washiyama}
\affiliation{GANIL, Bd Henri Becquerel, BP 55027, 14076 Caen Cedex 5, France}

\author{Denis Lacroix}
\affiliation{GANIL, Bd Henri Becquerel, BP 55027, 14076 Caen Cedex 5, France}

\author{Sakir Ayik}
\affiliation{Physics Department, Tennessee Technological University, Cookeville, TN 38505, USA}

\date{\today}

\begin{abstract}
Information on dissipation in the entrance channel of heavy-ion collisions is extracted 
by macroscopic reduction procedure of Time-Dependent Hartree-Fock theory. 
The method gives access to a fully microscopic description of the friction coefficient 
associated with transfer of energy from the relative motion towards intrinsic degrees of freedom. 
The reduced friction coefficient exhibits a universal behavior, i.e. almost 
independent of systems investigated, whose order of magnitude 
is comparable with the calculations based on linear response theory. 
Similarly to nucleus-nucleus potential, especially close to the Coulomb barrier, there are
sizable dynamical effects on the magnitude and form factor of friction coefficient. 

\end{abstract}

\maketitle

\section{Introduction}

The discovery of deep inelastic collisions in the 1970s,
in which a large amount of kinetic energy and angular momentum is dissipated
from the relative motion to intrinsic excitations of colliding nuclei,
brought us the concept of friction in nuclear physics~\cite{schroder84}. 
After this discovery, Gross and Kalinowski developed the surface friction model 
(SFM)~\cite{gross74} that takes into account this dissipative effect 
by introducing friction forces in a Newtonian equation.
More generally, dissipative aspects in self-bound system such as nuclei 
have important role on many physical phenomena such as  
fission, fusion reactions and giant resonances.

In most actual models that include dissipation and that are used to describe nuclear reactions
at energies around the Coulomb barrier,
dissipation mechanism is assumed to be a one-body type,
where energy dissipation is caused by collision of nucleons 
with mean-field wall and by nucleon exchange between the two partners of 
the reaction. These effects are known as the wall-and-window formula~\cite{blocki78,randrup80}
and are essentially based on classical consideration. The order of magnitude 
of parameters related to dissipation are generally adjusted 
to reproduce experiments. However, large uncertainty on those parameters exists~\cite{Hil93}.
Therefore, description of energy dissipation from quantum microscopic models is highly desirable.

It is know that time-dependent Hartree-Fock (TDHF) model~\cite{koonin80,negele82} 
includes one-body  dissipation mechanism
from  microscopic point of view, 
because of the treatment of self-consistent mean-field.
It is worth mentioning that so-called fusion window problem
due to underestimation of energy dissipation in old TDHF calculations
has been solved by including spin-orbit interactions 
and time-odd terms in the energy density functional~\cite{umar86,reinhard88,maruhn06}
as well as by breaking symmetries.
Now three-dimensional TDHF calculations 
including full Skyrme effective interaction, that are used in 
recent static Hartree-Fock calculations, are 
expected to provide a better description of dissipative 
aspects~\cite{kim97,nakatsukasa05,umar06,maruhn06}.
Up to now, there are only a few studies dedicated to extracting 
friction coefficient associated with one-body energy dissipation
from microscopic mean-field approach~\cite{koonin80,brink81}. 
The aim of this article is to investigate energy dissipation mechanism 
in detail from microscopic point of view by using state of art 
TDHF theory.

In Ref.~\cite{denis02,washiyama08}, a method of extracting nucleus-nucleus potential
and friction coefficient from TDHF has been proposed. This method, called Dissipative Dynamics TDHF (DD-TDHF), 
assumes that mean-field evolution can be properly reduced to one-dimensional dissipative 
dynamics for relative distance $R$ between nuclei. In Ref. \cite{washiyama08}, nucleus-nucleus 
potential for symmetric and asymmetric reactions has been systematically extracted from the DD-TDHF.
In this work, we focus on friction coefficient, which also comes as an 
output of the macroscopic reduction procedure. 

The paper is organized as follows. In section II, we briefly explain the DD-TDHF method. 
In section~III, we illustrate main properties  of extracted friction coefficient 
for different systems. In section~IV, we discuss a method to estimate intrinsic excitation of 
colliding nuclei, and compare the result to the dissipated energy.
A summary is given in section V.

\section{Description of the Dissipative-Dynamics TDHF method}

\begin{figure}[tbhp]
\includegraphics[width=0.9\linewidth, clip]{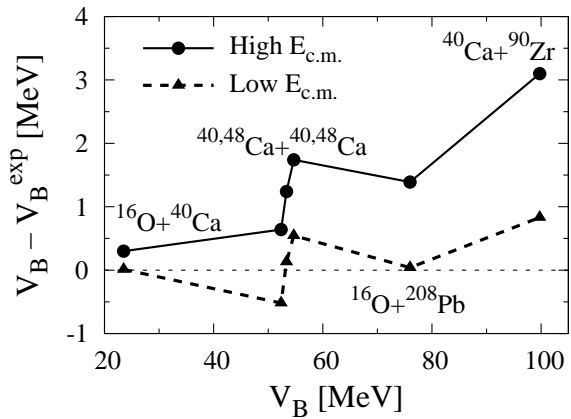}
\caption{
Barrier height extracted from DD-TDHF minus experimental barrier height 
as a function of extracted barrier height for reactions indicated in figure. 
$V_B$ is deduced from high c.m. energy (solid line) and from low c.m.  
energy TDHF (dashed line) reactions, respectively (Values of different barrier 
height can be found in table I of Ref.~\cite{washiyama08}).
}
\label{fig:systematics}
\end{figure}

Dissipative-Dynamics TDHF (DD-TDHF)  method, originally proposed in Ref.\cite{koonin80,denis02}, 
rely on the hypothesis that complex microscopic mean-field evolution of head-on collisions can
be accurately reduced into a simple one-dimensional macroscopic evolution given by,
\begin{eqnarray}
\frac{dR}{dt}&=&\frac{P}{\mu (R)},\\
\frac{dP}{dt}&=&-\frac{dV}{dR}-\frac{d}{dR}\left(\frac{P^2}{2\mu}\right)
                -\gamma (R)\dot{R}.
\label{newtonequation}
\end{eqnarray}
$R$ and $P$ denote here the relative distance and relative momentum between two nuclei, respectively, and 
are computed from mean-field theory according to the procedure described in Ref. \cite{denis02,washiyama08}. 
In the second equation, $V(R)$ and $\gamma (R)$ denote the nucleus-nucleus potential 
and friction coefficient, respectively, while term $-d_R(P^2/2\mu)$ 
arises from possible relative distance dependence of the reduced mass $\mu(R)$.  
The nucleus-nucleus potential for symmetric and asymmetric reactions 
deduced from DD-TDHF method has been systematically investigated in Ref. 
\cite{washiyama08}. Technical details of the method are extensively discussed 
in this reference and we recall here the main conclusions of the potential study: 
(i) We have shown in Ref.~\cite{washiyama08} that
the second term in the right hand side of Eq.~(\ref{newtonequation}),
i.e.,  term involving $R$-derivative of reduced mass, 
has a minor role on the extracted nucleus-nucleus potential and can be neglected. 
This is also the true here for relative distance larger than the barrier position, denoted by $R_B$.
In the following, we mainly focus on $R \ge R_B$, where the effect of the reduced mass $R$-dependence on 
friction coefficient is negligible.  
(ii) The DD-TDHF method gives similar results as the density-constrained TDHF (DC-TDHF) method 
at similar c.m. energy~\cite{Uma06}. 
(iii) As beam energy increase, the extracted potential tends towards the frozen density
(FD) approximation, which is also expected due to the fact that density has no time to reorganize as the 
nuclei collide. (iv) as c.m.  energy approaches the Coulomb barrier, dynamical effects 
induce a reduction of the barrier height, which is in agreement with fusion threshold 
deduced from TDHF in Ref.~\cite{Sim08} 
and is in close agreement with the experimental barrier height, see Fig.~\ref{fig:systematics}.
In this figure, the solid line corresponds to potential height extracted using
high c.m.  energies TDHF trajectories ($E_{\rm c.m.}\gg V_B$),
whereas the dashed line stands for barriers obtained when c.m.  energy used in 
DD-TDHF approaches the Coulomb barrier ($E_{\rm c.m.}\sim V_B$). 
Dynamical reduction of barrier height is clearly seen for all reactions. 
Due to this reduction, estimated barrier height at low energy becomes much closer 
to barriers extracted from experiments.
In summary, investigations of Ref. \cite{washiyama08} have shown  the simple macroscopic
reduction method of  DD-TDHF can provide a 
useful tool to infer nucleus-nucleus potentials from microscopic mean-field theory. 
Not only the potential, but employing the same method, we can extract information
about one-body dissipation mechanism, which is the topic of this paper. 

For TDHF calculations, we use
three-dimensional TDHF code developed by P.~Bonche and coworkers 
with the SLy4d Skyrme effective force~\cite{kim97}.
The mesh sizes in space and time are 0.8~fm and 0.45~fm/$c$,
respectively.
As TDHF initial conditions, 
we solve static HF equations~\cite{bonche85,ev8} 
with the same effective force and the same mesh size as in TDHF.
The initial distances between two nuclei are set 
between 16 and 22.4~fm.
We assume that colliding nuclei follow the Rutherford trajectory
before they reach the initial distance for TDHF calculations.

\section{Friction Coefficients from mean-field dynamics}

\subsection{Friction coefficient at c.m. energy well above the Coulomb barrier}

Besides nucleus-nucleus potential, one-dimensional 
macroscopic reduction gives also 
access to the friction coefficient $\gamma(R)$. We first focus on 
friction coefficients extracted from TDHF calculation when c.m. energy
is well above the Coulomb barrier, i.e. for which extracted potentials identify 
with FD potentials~\cite{washiyama08}. Possible c.m. energy dependence 
of dissipation will be discuss later.

\begin{figure}[tbhp]
\begin{center}\leavevmode
\includegraphics[width=0.9\linewidth, clip]{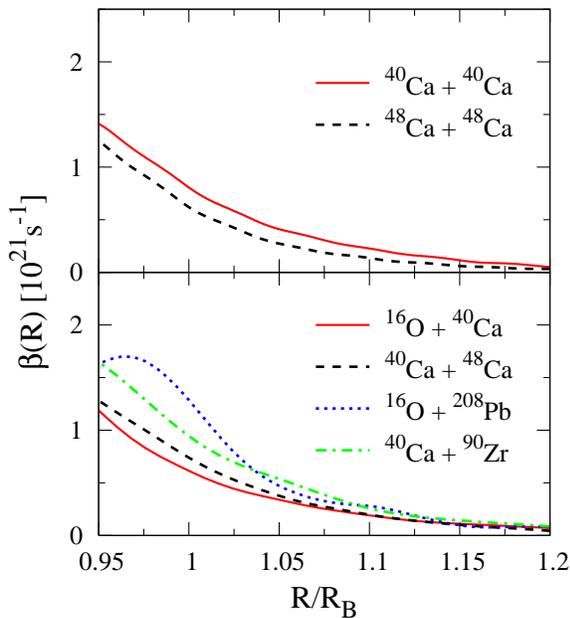}
\caption{
Reduced friction coefficient $\beta(R)=\gamma(R)/\mu(R)$ 
as a function of $R$ divided by the Coulomb barrier radius $R_B$
for mass symmetric (upper panel) and 
mass asymmetric (lower panel) reactions.
}
 \label{fig:frictionall}
\end{center}
\end{figure}
Fig. ~\ref{fig:frictionall}, presents 
an important results of this work. The reduced friction 
coefficients, defined as $\beta (R)=\gamma (R)/\mu (R)$, are 
extracted from a fully microscopic theory without any free 
parameter adjustment or adiabatic/diabatic assumption. 
Reduced friction coefficient is systematically shown 
for the mass symmetric reactions (upper panel) 
$^{40}$Ca${}+^{40}$Ca, $^{48}$Ca${}+^{48}$Ca  and
mass asymmetric reactions (lower panel) $^{16}$O$ +^{40,48}$Ca, 
$^{16}$O$+^{208}$Pb, 
$^{40}$Ca${}+^{48}$Ca, $^{40}$Ca$+^{90}$Zr, as a function of relative distance. 
As the colliding nuclei approach toward each other, reduced friction coefficient
monotonically increases and is almost independent of colliding 
system. Therefore, a universal behavior of the 
friction coefficient is observed in collisions at c.m.
energies well above the Coulomb barrier.

\subsection{Energy dependence of friction coefficient}

\begin{figure}[tbhp]
\begin{center}\leavevmode
\includegraphics[width=0.9\linewidth, clip]{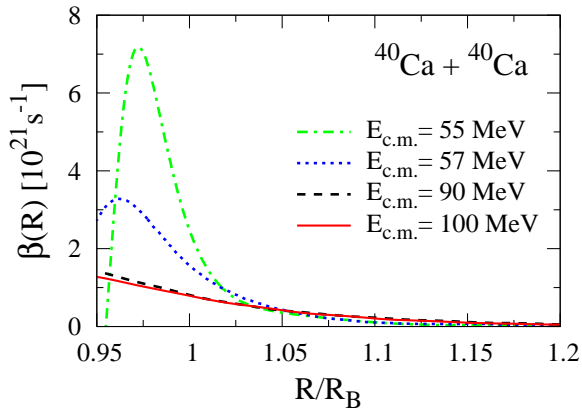}
\caption{Reduced friction coefficient 
as a function of $R/R_B$ for the $^{40}$Ca$+^{40}$Ca reaction
at different c.m. energies.
}
 \label{fig:caca}
\end{center}
\end{figure}
As c.m.  energy approaches the Coulomb barrier, 
dynamical effects such as deformation and neck formation 
can take place. These aspects are automatically included in 
dynamical mean-field calculation and have been shown to 
systematically reduce the barrier compared to the high energy limit.
Similarly to potential energy landscape, we do expect
dissipation to be modified as the beam energy decreases. 
Fig.~\ref{fig:caca} shows the extracted friction coefficient
obtained using different c.m. energies between 55 and 100~MeV 
for the $^{40}$Ca+$^{40}$Ca reaction. 
We note that the Coulomb barrier energy of this reaction 
extracted at c.m. energy $E_{\rm c.m.}=55$~MeV is about 53~MeV.
Similarly to potential landscape, which tends to the FD case as
the beam energy increases, we observe that the magnitude of
friction coefficient 
does not change as the c.m. energy increases between 
$E_{\rm c.m.}=90$~MeV and 100 MeV. The associated dissipation 
corresponds to limited situations 
when density has no time to reorganize in the entrance channel.

On the other hand, for $E_{\rm c.m.}=55$~and 57~MeV,
which are close to the Coulomb barrier,
friction coefficient exhibits sizable energy dependence. 
The lower the energy is the larger the magnitude of friction.  
Radial dependence of friction coefficient is 
very different from that at high energies. At low energies,
friction coefficient shows a peak near the Coulomb barrier 
and decreases to smaller values inside the Coulomb barrier,
which is located around $R=R_B\approx10$~fm.

\begin{figure}[tbhp]
\begin{center}\leavevmode
\includegraphics[width=0.9\linewidth, clip]{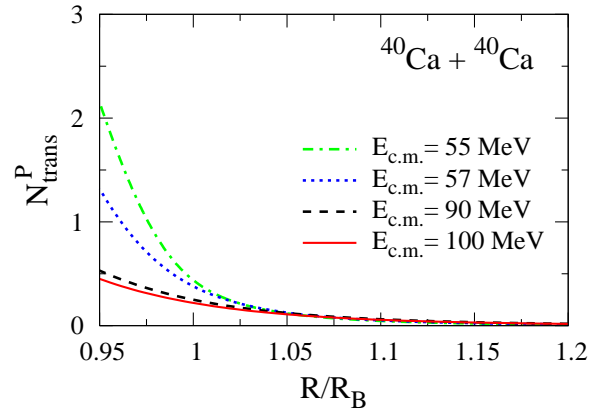}
\caption{Number of nucleon transferred from projectile to target $N^{P}_{\rm trans}[R(t)]$  
as a function of $R$ for the $^{40}$Ca$+^{40}$Ca reaction
at the same c.m. energies as in Fig.~\ref{fig:caca}.
}
 \label{fig:transfer}
\end{center}
\end{figure}

The enhanced dissipation around the Coulomb barrier energies is partly due to 
early neck formation accompanied by an increase of nucleon exchange. This is 
illustrated in Fig.~\ref{fig:transfer}, where the number of nucleon
transferred from one nucleus to the other is shown as a function of $R$ for the $^{40}$Ca$+^{40}$Ca reaction
at same energies as in Fig.~\ref{fig:caca}. For such a symmetric reaction, the separation plane 
is located at $x=0$. The number of nucleon initially in the projectile (taken by convention 
at initial time at $x<0$)
having passed through the separation plane, i.e., transferred to the target, is estimated by,  
\begin{equation}
N^{P}_{\rm trans}[R(t)] = \int d^3x \,\rho_P({\mathbf r},t)\,\theta(x),
\end{equation}
where $\rho_P$ (resp. $\rho_T$) are defined through
\begin{eqnarray}
\rho_{P/T}({\mathbf r},t)=\sum_{i \in {P/T}} |\phi_i({\mathbf r},t)|^2,
\label{eq:rhopt}
\end{eqnarray} 
with $\phi_i({\mathbf r},t)$ being the single-particle
wave function initially in the projectile (resp. target) 
and propagated through the mean-field. $\theta(x)$ is the
step function equal to zero for negative $x$ and $1$ for positive $x$. We can equivalently define 
number of nucleons transferred from target to projectile, denoted by $N^T_{\rm trans}$. 
For symmetric reactions, $N^P_{\rm trans}$ is equal to $N^T_{\rm trans}$ at all time.
For $R \ge R_B$, enhancement of dissipation observed in Fig.~\ref{fig:caca} is strongly correlated to  
increase of particle exchange. 
Such a strong correlation is indeed expected 
from the window dissipation mechanism~\cite{randrup80}, in which
the magnitude of friction coefficient is proportional to 
the window area, and hence the number of nucleons exchanged through the window.
From strong similarity between Figs.~\ref{fig:caca} and \ref{fig:transfer}, one 
can conclude that the main source of dissipation at large distance ($R \ge R_B$) 
is due to nucleon exchange.  

\begin{figure}[tbhp]
\begin{center}\leavevmode
\includegraphics[width=0.9\linewidth, clip]{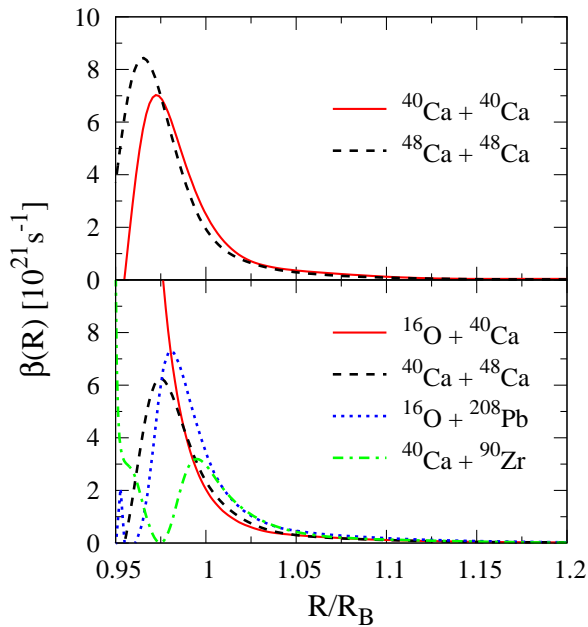}
\caption{
Reduced friction coefficient $\beta(R)=\gamma(R)/\mu(R)$ 
as a function of $R$ divided by Coulomb barrier radius $R_B$
for mass symmetric (upper panel) and 
mass asymmetric (lower panel) reactions. In each case, 
c.m. energy used to extract friction coefficient corresponds 
to the Coulomb barrier energy.
}
 \label{fig:frictionalllow}
\end{center}
\end{figure}

Enhancement of dissipation at low c.m. energy is systematically observed
as seen in Fig.~\ref{fig:frictionalllow}, where the 
reduced friction coefficient $\beta(R)$ is plotted as a function of relative distance. 
It also appears that enhancement of dissipation before the barrier $R \ge R_B$
is nearly independent of systems investigated. For instance near barrier $R/R_B = 1$, 
the magnitude of friction at high energy is about $\beta(R) \equiv 0.8 \times 10^{21}s^{-1}$,
while at near barrier energies, its magnitude is about four times larger 
$\beta(R) \equiv 3 \times 10^{21}s^{-1}$, for all systems investigated. 
 
For all the systems investigated, 
calculations show that the magnitude of friction coefficient rapidly
decreases for decreasing relative distances $R < R_B$. We believe that this
unphysical behavior is due to simple macroscopic reduction
procedure, which breaks down at low energies, when colliding nuclei begins to overlap strongly.
In this case, complex dynamical effects such as onset of nuclear 
deformation~\cite{washiyama08} and non-Markovian effects become important in dissipation
mechanism, which are not incorporated in the simple reduction procedure presented here.

\subsection{Comparison with other models}

There is large uncertainty between microscopic and phenomenological description of
nuclear dissipation~\cite{Hil93}. Therefore it is of great interest to provide an
accurate description of nuclear dissipation. The macroscopic reduction presented in this work
serves a useful insight towards that goal. In this section, we compare our results
with macroscopic surface friction model (SFM)~\cite{gross74} and microscopic
calculations of Adamian et al.~\cite{adamian97}.
The SFM was introduced in a classical description of relative motion in deep-inelastic
heavy-ion collisions in Ref.~\cite{gross74}. The radial friction force was parameterized as 
$\gamma(R)=K_r(dV_{N}/dR)^\alpha$, 
where $V_{N}$ is the nuclear part of the nucleus-nucleus potential 
and $K_r$ and $\alpha$ are parameters.
These parameters were fitted as $K_r=4\times 10^{-23}$~s/MeV and $\alpha = 2$. 
Microscopic calculations of Ref. \cite{adamian97} is based on the linear response
theory of nuclear dissipation~\cite{hofmann76}. These calculations
take into account time evolution of 
single-particle occupation factors during collision through 
a consistent treatment of the collective and intrinsic degrees of freedom. 

\begin{figure}[tbhp]
\begin{center}\leavevmode
\includegraphics[width=0.9\linewidth, clip]{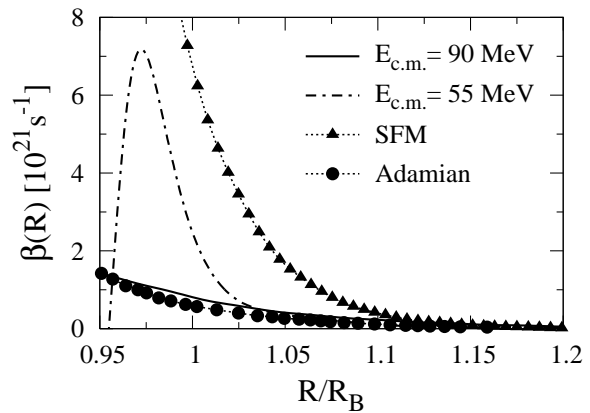}
\caption{Comparison between reduced friction coefficient deduced with DD-TDHF 
method for the $^{40}$Ca$+^{40}$Ca reaction at $E_{\rm c.m.}=90$~MeV (solid line)
and the one computed by Adamian et al.~\cite{adamian97} for 
the $^{64}$Zn$+^{196}$Pt reaction at $E_{\rm lab}=440$~MeV (filled square).
The result of SFM~\cite{gross74} for the $^{40}$Ca$+^{40}$Ca reaction 
is also presented by filled triangles.
}
\label{fig:compare}
\end{center}
\end{figure}
In Fig.~\ref{fig:compare}, high energy (solid line) and low energy (dot-dashed line) DD-TDHF results 
are compared both with the SFM case (filled triangles) and 
an example of linear response theory result obtained in Ref.~\cite{adamian97} (filled circles).
It is observed that at $R \approx R_B$ the SFM strongly overestimates 
the magnitude of friction coefficient compared to the other 
microscopic models, while our result agrees at high c.m.  energy 
with the microscopic calculations 
of Ref.~\cite{adamian97} (filled squares).

\section{Dissipated energy, nucleon exchange and excitation energy in nuclei}

In this section, we discuss the link between  internal excitation energy 
of colliding nuclei and dissipated energy from the macroscopic degrees of freedom.
In addition, we give further evidence for the fact that particle transfer is the main source 
of dissipation before two nuclei reach the Coulomb barrier.  
\subsection{Total Energy dissipation}
\begin{figure}[tbhp]
\begin{center}\leavevmode
\includegraphics[width=0.9\linewidth, clip]{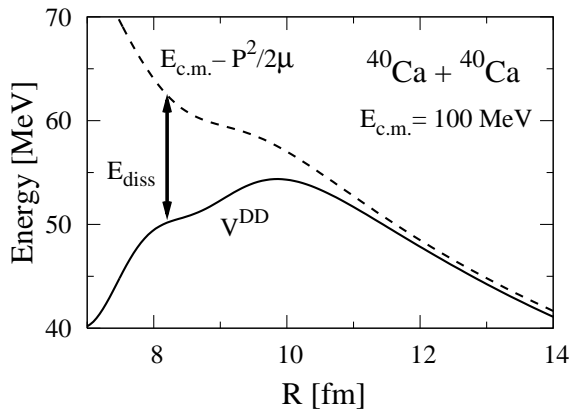}
\caption{Evolution of potential $V^{DD}$ extracted from the DD-TDHF
as a function of $R$ (solid line) for the $^{40}$Ca{}+$^{40}$Ca reaction at $E_{\rm c.m.} = 100$~MeV. 
The difference between  total c.m. energy and relative kinetic energy is also shown by dashed 
line. Difference between  two curves is nothing but dissipated energy in the entrance channel.
}
\label{fig:ediss1}
\end{center}
\end{figure}

From energy conservation, we can give a simple estimate of dissipated energy
$E_{\rm diss}$ in the entrance channel as,
\begin{eqnarray}
E_{\rm diss} &=& E_{\rm c.m.} - \frac{P^{2}}{2 \mu} -V^{DD}(R),
\label{eq:ediss0}
\end{eqnarray}
where the second term is nothing but the relative kinetic energy while $V^{DD}$ denotes the 
potential extracted from the DD-TDHF method.  
Fig.~\ref{fig:ediss1} illustrates magnitude of different quantities as a function
of relative distance for the $^{40}$Ca{}+$^{40}$Ca reaction at $E_{\rm c.m.} = 100$~MeV.
According to Eq.~(\ref{newtonequation}), we can calculate dissipated energy from 
the friction coefficient $\gamma$ using the Rayleigh formula:  
\begin{equation}
E_{\rm diss}(R(t))=\int_0^t dt' \,\gamma(R(t'))\,[{\dot R(t')}]^2,
\label{eq:ediss}
\end{equation}
where $\dot{R}$ denotes the relative velocity deduced from mean-field evolution. We have checked that the 
above equation gives identical result as Eq.~(\ref{eq:ediss0}). 

The dissipated energy provides a measure for transfer of
energy from  relative motion to internal degrees of 
freedom during the collision. Denoting by $E^*$ the internal excitation energy and assuming that all dissipated 
energy is converted into internal excitation energy, we do expect $E^* = E_{\rm diss}$. In the
following, we apply a method to estimate directly the internal excitation energy of  
projectile and target during the early stages of collision and show that a good agreement is obtained with the 
amount of dissipated energy.

\subsection{Estimate of internal excitation energy}

\subsubsection{Single nucleus case}

For a single isolated nucleus, in the mean-field approach, 
the excitation energy can be estimated using 
\begin{eqnarray}
E^* = {\cal E}^{ex}_{MF} - {\cal E}^0_{MF}, 
\end{eqnarray} 
where ${\cal E}^0_{MF}$ and ${\cal E}^{ex}_{MF}$ denote
ground state and excited state mean-field energy, respectively.
For small excitations, we can 
approximately calculate the excitation energy according to,
\begin{eqnarray}
E^* \simeq \sum_i (\varepsilon^0_i- \varepsilon^\tau_F) \times (n_i -n^0_i). 
\label{estarone}
\end{eqnarray} 
In this expression, $\varepsilon^0_i$ denotes single-particle energies,
$\varepsilon^\tau_F$, with $\tau=n,p$ stands for the neutron or proton Fermi energy,  
while $n_i$ and $n^0_i$ are the occupation factors in the excited 
and the ground state, respectively. 
\subsubsection{Dinucleus case}
\begin{figure}[tbhp]
\begin{center}\leavevmode
\includegraphics[width=0.9\linewidth, clip]{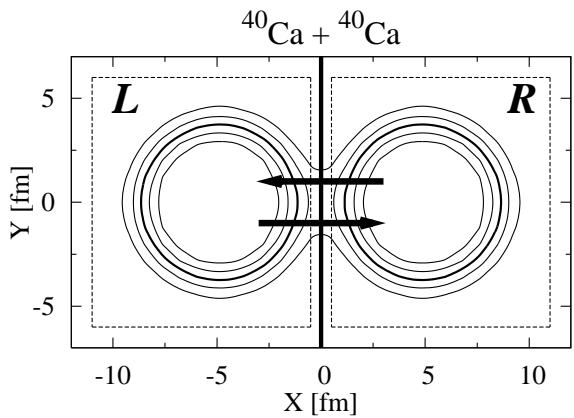}
\caption{Schematic illustration of the $^{40}$Ca{}+$^{40}$Ca reaction in
the entrance channel at relative distance $R=9.8$~fm.  
Different curves correspond to iso-contours of total nuclear density
while vertical line denotes the separation plane. 
}
 \label{fig:schemsep}
\end{center}
\end{figure}

\begin{figure}[tbhp]
\begin{center}\leavevmode
\includegraphics[width=0.9\linewidth, clip]{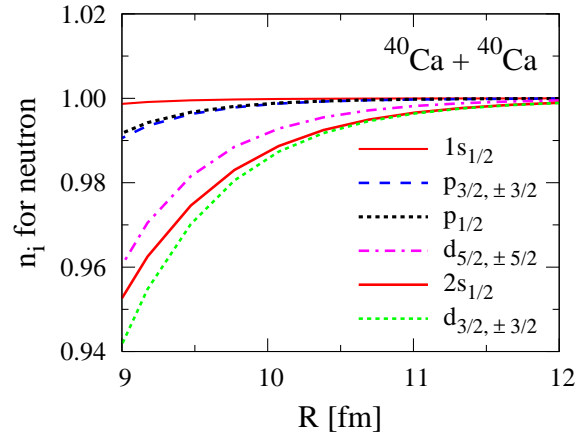}
\caption{
Occupation factors $n_i=\langle i|i\rangle_L$ for neutron single-particle states as 
a function of $R$ for the $^{40}$Ca$+^{40}$Ca reaction at $E_{\rm c.m.}=100$~MeV.
}
\label{fig:occupation}
\end{center}
\end{figure}

In the case of two colliding nuclei, we are mainly interested in the entrance channel
($R \ge R_B$) where the two nuclei slightly overlaps. In this di-nuclear configuration, 
we define the collective variables by drawing the separation plane as illustrated
in Fig.~\ref{fig:schemsep}.  
In the entrance channel, colliding nuclei are weakly excited. In order to
calculate the excitation energy of each partner using the lowest order 
perturbation expression (\ref{estarone}), we need the occupation factors of single-particle 
states. We can determine the occupation factors by constructing the overlap matrix
of time-dependent single-particle states in the right side of the 
separation plane~\cite{dasso79},
\begin{equation}
 \langle i|j\rangle_{R}=\int d^3r \,\phi_i^*({\bf r},t)\,\phi_j({\bf r},t)\,\theta(x-x_0),
\end{equation}
where the separation plane is at $x=x_0$. In this expression, 
$\phi_i({\bf r},t)$ denote 
single particle states originating either from the target or projectile. The
overlap matrix in the left side $\langle i|j\rangle_{L}$ is defined similarly.
Occupation factors $n_\alpha$ associated to the left and right sides on the
separation plane are obtained by diagonalizing the corresponding overlap matrix.
At large relative distance, the occupation factors can be grouped into two classes. 
We first consider the left sub-system which initially contains the projectile.
States with eigenvalues of $\langle i|j\rangle_{L}$ close to one will correspond 
to single-particle states originating from projectile side (left side), 
while those with eigenvalues close to 
$0$ correspond to single-particle states originating  from the target 
(right side) and which are penetrated to the left. Since in the entrance channel,
changes of occupation factors are small, instead of carrying out diagonalization,
we can use first order perturbation theory to determine  occupation factors.
In first order perturbation theory, eigenvalues of the overlap matrices 
$\langle i|j\rangle_{L}$ are given by $n_\alpha\approx\langle i|i\rangle_L \equiv n_i$, 
where $|i\rangle$ denotes a complete basis of the projectile 
including initially unoccupied states. 
As an example, Fig.~\ref{fig:occupation} shows evolution of occupation factors $n_i$
for six neutron single-particle states initially corresponding to
$1s_{1/2}$, $p_{3/2,\pm 3/2}$, $p_{1/2}$, $d_{5/2,\pm 5/2}$, $2s_{1/2}$, and $d_{3/2,\pm 3/2}$ states
(with the notation $\ell_{j,j_z}$) as a function of $R$
for the $^{40}$Ca$+^{40}$Ca reaction at $E_{\rm c.m.}=100$~MeV. 
Occupation factors decrease monotonically as $R$ decreases but remains still close to one
around the Coulomb barrier $R_B\approx 9.8$~fm. 

In order to have a simple estimate of the excitation energy, we further assume that 
occupation factors $n_i=\langle i|i\rangle_L$ have the shape as the Fermi-Dirac distribution.
Then, only tail of the distribution contributes to the excitation energy, consequently, the
excitation energy of the left sub-system can approximately be calculated according to
\begin{eqnarray}
E_L^{*}(t)&\approx& 2 \sum_{i=1}^{A_P}(\varepsilon^0_i -\varepsilon^\tau_F) (\langle i|i\rangle_L - n^0_i),
\label{eq:el}
\end{eqnarray}
where summation runs over  
the states that are initially occupied in the projectile.
Similar expression can be found for the excitation $E^{*}_R$ of the right sub-system. 
\begin{figure}[tbhp]
\begin{center}\leavevmode
\includegraphics[width=0.85\linewidth, clip]{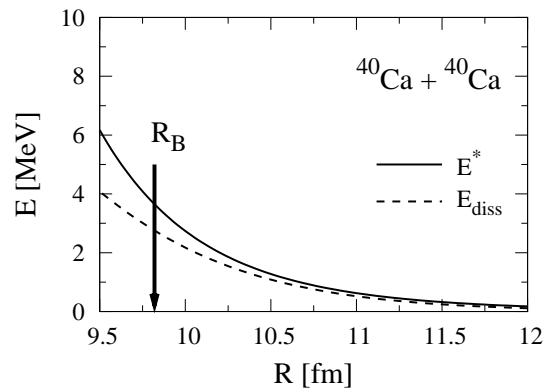}
\caption{
Comparison of internal excitation energy $E^*$ (solid line)
with dissipated energy $E_{\rm diss}$  from Eq.~(\ref{eq:ediss}) (dashed line) 
as a function of $R$ for the $^{40}$Ca$+^{40}$Ca reaction at $E_{\rm c.m.}=100$~MeV.
Vertical arrow indicates the position of the Coulomb barrier radius. }
 \label{fig:excitation}
\end{center}
\end{figure}
In Fig.~\ref{fig:excitation}, total excitation energy $E^*=E_R^{*}+E_L^{*}$ (solid line) 
is compared with dissipated 
energy $E_{\rm diss}$ given by Eq. (\ref{eq:ediss}) (dashed line)
for the $^{40}$Ca$+^{40}$Ca reaction at $E_{\rm c.m.}=100$~MeV. The agreement between 
two quantities is quite good all the way up to the barrier distance $R>R_B$. 
Near equality of the excitation energy and the dissipated energy provides a 
consistency check the friction coefficient extracted from the TDHF simulations.

\section{summary}

Using a macroscopic reduction procedure of the mean-field theory
proposed in Ref.~\cite{koonin80,denis02,washiyama08},
we extract the friction coefficient associated with one-body 
energy dissipation in the entrance channel of heavy-ion fusion reactions.
The magnitude and form factor of reduced friction coefficient have a universal 
property for various reactions which are investigated. 
Nucleus-nucleus potentials obtained
with the same method exhibits energy dependence. In a similar manner, 
magnitude and form factor of the extracted friction coefficient depends  
on the beam energy as well. It is observed that the rate of dissipation 
increases as beam energy 
approaches the Coulomb barrier. The enhancement of dissipation rate at low energy 
is a consequence of early neck formation and increasing rate of particle exchange
between projectile-like and target-like nuclei. 
The order of magnitude of dissipation 
deduced from TDHF is in agreement with microscopic calculations of friction 
coefficient based on the linear response theory. 
We estimate the excitation energy in the entrance channel by other method.
Very close agreement found between the calculated excitation energy and the
dissipated energy determined from the friction coefficient provides a further
support for the validity of the extracting procedure for dissipation that 
we employed.

\begin{acknowledgments}
We thank P. Bonche for providing us the 3D-TDHF code. We also are grateful to 
B. Avez, D. Boilley and C. Simenel for discussions. One of us (S.A.) gratefully
acknowledges CNRS for financial support and GANIL for warm hospitality extended
to him during his visit. This work is supported in part by the US DOE Grant No.
DE-FG05-89ER40530.

\end{acknowledgments}

\end{document}